\documentclass[twocolumn,showpacs,showkeys,preprintnumbers,amsmath,amssymb,aps]{revtex4}
\usepackage{graphicx}
\usepackage{float}
\usepackage{amssymb}

\begin{document}

\preprint{Giant magnetoresistance in semiconducting DyNiBi, Casper \it{et al}}
\title{Giant magnetoresistance in semiconducting DyNiBi}

\author{Frederick~Casper,  and Claudia~Felser}
\email{casperf@uni-mainz.de}
\affiliation{Institut f\"ur Anorganische Chemie und Analytische Chemie,\\
Johannes Gutenberg - Universit\"at, 55099 Mainz, Germany.\\}

\begin{abstract}
The semiconducting half-Heulser compound DyNiBi shows a negative giant magnetoresistance (GMR) below 200 K. Except for a weak deviation, this magnetoresistance scales roughly with the square of the magnetization in the paramagnetic state, and is related to the metal-insulator transition. At low temperature, a positive magnetoresistance is found, which can be suppressed by high fields. The magnitude of the positive magnetoresistance changes slightly with the amount of impurity phase.    

\end{abstract}



\maketitle
\section{Introduction}
Magnetoresistive sensors for information storage play an important role in the information technology. At the moment, the fundamental physical limits of the current technology have been reached, namely the size of the structure. Within the last decade, a new approach using the spin of electrons instead of the charges has been developed \cite{FEL07}. Devices that have been built utilizing magnetoresistive (MR) effects have become important, e.g., devices built upon the giant magnetoresistance (GMR) effect \cite{GRU89,FER88}. Tunnel magnetoresistance (TMR) was exploited in constructing magnetic tunnel junctions (MTJs); MTJs based on full-Heusler compounds seem to be especially good candidates for building devices \cite{TEZ06}.  

This letter reports on the GMR in the semiconducting bulk compound DyNiBi. In the semiconducting ternary rare earth compounds {\it RE}NiSb (RE = Tb, Dy, Ho), which have the same half-Heusler structure as DyNiBi, a large magnetoresistance has been found at low temperature \cite{KAR98}. The magnetoresistance of these compounds have been studied extensively \cite{PIE99,PIE00}. Magnetoresistance often appears in narrow gap semiconductors having an internal magnetic shell, thus one can expect large magnetoresistance in the {\it RE}NiBi series, which have a smaller gap than the {\it RE}NiSb compounds \cite{LAR99}. Recently, it was shown that 18 valence electron compounds with this structure (MgAgAs structure) are closed shell species, non-magnetic and semiconducting \cite{KAN06}. The {\it f}-electrons of the rare earth metal are localized and therefore not considered as valence electrons. 

At low temperatures, a positive contribution to the magnetoresistance is found, similar to other rare earth containing compounds (e.g. \cite{SEN04}) and doped ZnO films (e.g. \cite{GAG07}). On the other hand, the magnitude of the positive MR slightly increases with higher impurity contend.   

\section{Experimental Details}
DyNiBi samples were prepared by arc melting the pure elements under a dry argon atmosphere. A slight excess of bismuth (from 5 up to 20$\%$ for different samples) was used to compensate for losses during the arcmelting process. Nevertheless, the melting was carried out with low current to minimize the loss of bismuth. To enhance homogeneity, the samples were turned over at least once. The resulting ingots were wrapped in tantalum foil, sealed under vacuum into silica tubes and annealed for 7 day at 750 °C.
      
The structure was investigated by means of x-ray diffraction (XRD) using Cu - K$_\alpha$ radiation.  DyNiBi crystallizes in the cubic MgAgAs structure type (C1$_b$ structure, {\it F}$\bar 4$3{\it m}) with {\it a} = 0.64108~nm, Dy at 4{\it b} ($\frac{1}{2}\: \frac{1}{2}\: \frac{1}{2}$), Ni at 4{\it c} ($\frac{1}{4}\: \frac{1}{4}\: \frac{1}{4}$) and Bi at 4{\it a} (0~0~0), which is in good agreement with \cite{DWI74}. Magnetic measurements show an antiferromagnetic transition at T$_N=$~3~K.

\begin{figure}[H]
\centering
\includegraphics[width=7.5cm]{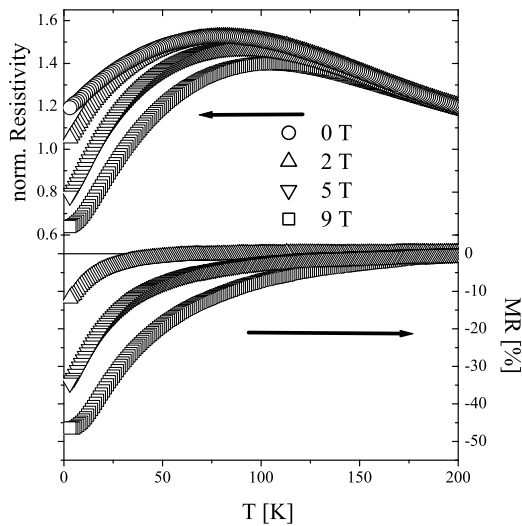}
\caption{Resistivity and magnetoresistance of DyNiBi in different magnetic fields. The resistivity has been normalized to value of 300 K/ zero field.}
\label{Resis}
\end{figure}

Figure~\ref{Resis} shows the temperature dependence of the resistivity in different magnetic fields. The metal to semiconductor transition temperature strongly depends on the degree of order in the sample. The C1$_b$ structure can be thought as Heusler structure (L2$_1$ structure) with a vacant site. Thus Ni atoms can occupy the vacant site; antiphase domains corresponding to the inversion of Dy and Bi may occur. The degree of order and thus the transition temperature can be changed by simple annealing of the sample. Longer annealing enhances the degree of order and shifts the transition to lower temperatures (not shown here). 
For the temperature range 200 - 300K, the resistivity $\rho_a$ follows the activation law:

\begin{equation} 
   \rho_a = \rho_0 \exp(E_a/k_BT),
\label{eq1}
\end{equation}

 where $E_a \approx$~10~meV. For temperatures lower than 100 K a metallic-like behavior is observed. Such a behavior is typical for conduction through an impurity band that overlaps the conduction band. The resistivity $\rho_m$ can be approximated by: 

\begin{equation} 
   \rho_m = \rho_0 + CT^\alpha , 
\label{eq2}
\end{equation}

with $\alpha \approx$~0.6. Thus the complete temperature dependence can be described by the conduction of parallel channels, one corresponding to the metallic conduction and the other to the thermal activation. The temperature dependence of the MR is weak above 150 K, but increases at lower temperatures. The resistivity slope shows a metal - insulator (MI) transition. This transition strongly depends on the preparation conditions, e.g., longer annealing of the samples shifts the transition to lower temperatures.  

Non-magnetic rare earth compounds have a low positive magnetoresistance. This weak MR may arise from the additional scattering of the charge carriers due to the Lorenz force (see e.g. \cite{ZIM63}). For similar rare earth containing semiconductors with a magnetic transition, a negative MR is found that, for low MR values in the paramagnetic state \cite{PIE00}, is proportional to the square of the magnetization of the 4{\it f} shell and is given by: \\ 

\begin{equation} 
	MR = -AM^2(H,T),
\label{eq3}
\end{equation}
 
\begin{figure}[H]
\centering
\includegraphics[width=7.5cm]{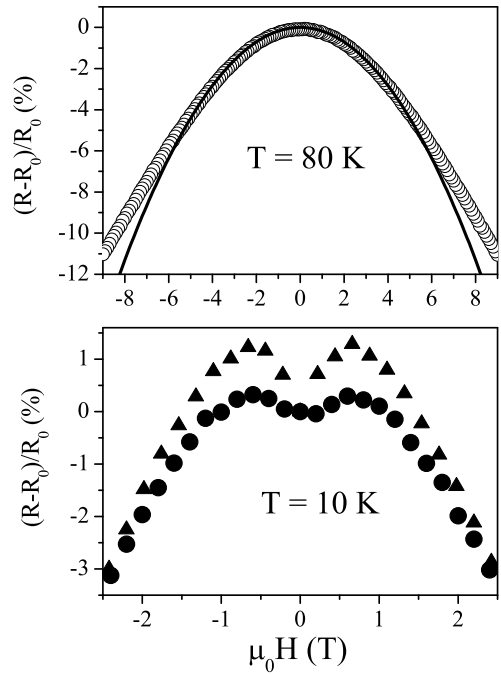}
\caption{Field dependence or the resistvity at 80 K (top) and 10K (bottom). The straight line is the fit according to equation~\ref{eq3}. At 10 K two samples with slightly different bismuth impurities were measured (triangles = higher impurity contend).}
\label{fig3_mr}
\end{figure}

Figure~\ref{fig3_mr} shows the field dependence of the resistivity at 80 K (open circles). A fit to the experimental data (solid line) shows the {\it M$^2$} variation of the MR with {\it A} = 1.767 $\times$ 10$^{-9}$ kOe$^{-2}$. At 10 K, the sign of the MR changes in low fields. The magnitude of the positive MR is slightly different for the measured different samples. The compound with the higher impurity contend shows a slightly higher positive MR.     

\begin{figure}[H]
\centering
\includegraphics[width=7.5cm]{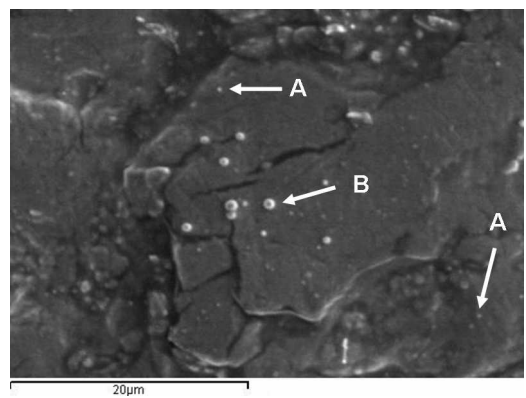}
\caption{Scanning microscope image of a polished DyNiBi surface. Arrows mark Bi dots in the DyNiBi matrix (A) and on the surface (B) due to soft polishing}
\label{fig1_rem}
\end{figure}

Scanning electron microscope (SEM) measurements reveal the inhomogeneity of the sample (Figure~\ref{fig1_rem}). Small semimetallic Bi dots (A) are clearly visible in the semiconducting DyNiBi matrix. The Bi dots on the surface (B) are due to the soft polishing process. The size of the Bi dots is 400 - 800 nm.

The positive MR in low fields and at low temperatures can be explained in two different ways. First, it could be due to changes to the density of states and mobility for up- and down-spin carries as found in EuSe \cite{SHA74}. Second, the positive MR  could be caused by the metallic bismuth impurities. It is known that inhomogeneities have significant effects on the physical properties of semiconductors \cite{WEI63,HER60}, and, in particular on the magnetoresistance \cite{THI98}. The embedded bismuth acts as a short circuit with most of the applied current passing through it. When a magnetic field is applied, the semimetallic inhomogeneities act as open circuits and the total resistivity rises \cite{MOU01}. This geometric contribution to the GMR, the so-called extraordinary magnetoresistance (EMR) \cite {SOL00} constitutes a significant change of the MR at fields below 6 kOe. As the magnitude of the positive MR changes at different impurity contend of the samples (Figure~\ref{fig3_mr}b), the positive MR is more likely attributed to the EMR. In higher fields, the EMR is overwhelmed by the negative GMR. Similar to other semiconducting rare earth compounds, the negative GMR effect is due to the polarization of the current carriers by the 4 {\it f} shell \cite{PIE00}.   

The goal is to synthesis new materials for spin applications. The compound shows a negative GMR below 200 K. The transition depends strongly on the sample preparation conditions. A higher charge carrier density should shift the GMR to higher temperatures \cite{BAL07}. Possible due to small Bi inhomogeneities, the compound also shows a positive MR at low temperatures and fields whose origin can be explained by the extraordinary magnetoresistance effect (EMR). For these cases, the EMR is suppressed in higher fields by the negative MR of the DyNiBi phase. By tuning the bismuth impurities, the EMR effect is affected. Semiconductor-metal hybrid structures are considered to be of enormous technological interest for the development of, e.g., ultrafast read heads. Using this approach could lead to the development of a simple bulk sensor for magnetic fields. 

\bigskip

We thank S. E. Wolf (Mainz, Germany) for the help with the SEM, as well as D. Weiss (Regensburg, Germany) and Bruce Gurney (Hitachi Global Storage Technologies, San Jose, USA) for useful discussions. Financial support by DfG grant FE633/1-1 within SPP1166.


\end{document}